\documentclass{article}
\usepackage[T1]{fontenc}
\usepackage{geometry}
\usepackage{color}
\usepackage{setspace}
\makeatletter

\providecommand{\LyX}{L\kern-.1667em\lower.25em\hbox{Y}\kern-.125emX\@}
\newcommand{\noun}[1]{\textsc{#1}}

 \newcommand{\lyxaddress}[1]{
   \par {\raggedright #1 
   \vspace{1.4em}
   \noindent\par}
 }

\makeatother

\begin{document}

\title{\textcolor{black}{\huge Interpolating gauges and the importance of a careful
treatment of \( \epsilon  \)-term}\huge }

\author{Satish D. Joglekar\thanks{
e-mail address: sdj@iitk.ac.in
}}

\maketitle

\lyxaddress{\( \qquad \qquad \qquad  \)Department of Physics,  I.I.T.Kanpur , Kanpur 208016
{[}India{]}}

\begin{abstract}
We consider the use of interpolating gauges (with a gauge function \( F[A;\alpha ] \)
) in gauge theories to connect the results in a set of different gauges in the
path-integral formulation. We point out that the results for physical observables
are very sensitive to the epsilon term that we have to add to deal with singularities
and thus it cannot be left out of a discussion of gauge-independence generally.
We further point out, with reasons, that the fact that we can ignore this term
in the discussion of gauge independence while varying of the gauge parameter
in Lorentz-type covariant gauges is an exception rather than a rule . We show
that generally preserving gauge-independence as \( \alpha  \) is varied requires
that the \( \epsilon  \)-term \emph{has} to be varied with \( \alpha  \).
We further show that if we make a naive use of the (fixed) epsilon term \( -i\varepsilon \int d^{4}x[\frac{1}{2}A^{2}-\overline{c}c] \)
(that is appropriate for the Feynman gauge) for general interpolating gauges
with arbitrary parameter values {[}i.e. \( \alpha  \){]} , we cannot preserve
gauge independence {[}except when we happen to be in the infinitesimal neighborhood
of the Lorentz-type gauges{]}. We show with an explicit example that for such
a naive use of an \( \epsilon  \)-term, we develop serious pathological behavior
in the path-integral as \( \alpha  \) is/are varied. We point out that correct
way to fix the \( \epsilon  \)-term in a path-integral in a non-Lorentz gauge
is by connecting the path-integral to the Lorentz-gauge path-integral \emph{with
correct \( \epsilon  \)-term} as has been done using the finite field-dependent
BRS transformations in recent years.
\end{abstract}

\section{INTRODUCTION}

Calculations in the standard model, a non-abelian gauge theory, have been done
in a variety of gauges depending on the ease and convenience of calculations
{[}1{]}. Many different gauges have also been used in formal treatments in different
contexts. For example axial gauges have been used in the treatment of Chern-Simon
theory and Coulomb gauges in the confinement problem in QCD {[}2{]}. Superstring
theories also use to advantage both the covariant and the light-cone treatments
{[}3{]}. One of the important questions, far from obvious, has been whether
the results for physical observables are, in fact, independent of the choice
of the gauge used in calculations. While it has naively been assumed that this
must do so, it is quite another matter to actually prove the gauge independence
of observables in general\footnote{%
As the path-integral in Lorentz-type gauges are well-defined, a little thought
will show that in the path-integral framework this is really a question of whether
and how path-integrals in other sets of gauges can be defined in a manner consistent
with the Lorentz gauges.
}. A good deal of literature has been devoted to this question {[}4,5{]} directly
or indirectly. In fact, development of gauges other than Lorentz gauges has
not been an easy and completed task {[}2,4{]}. One of the methods commonly used
for this purpose in the literature is that of interpolating gauges {[}5,6,7{]}.The
basic idea behind interpolating gauges is to formulate the gauge theory in a
gauge for which the gauge function \( F[A(x);\alpha ] \) depends on one or
more parameters \( \alpha  \) in such a manner that for different values of
the parameters we recover gauge theories in different gauges. For example the
gauge function \( F[A(x);\theta ] \) used by Doust {[}7{]} to connect the Coulomb
and the Feynman gauge is given by

\( F[A,\theta ] \) = \( [\theta \partial ^{0}A_{0}-\frac{1}{\theta } \)\( \partial  \)\( _{i}A_{i}] \)
~ ~~ ~~ ~~ ~~ ~ ~~ ~~ ~~ ~~~ ~~ ~~~~~ ~~ ~~ ~~
~ ~~ ~~ ~~ ~~~(1.1)

where for \( \theta  \)=1, we recover the Feynman gauge and for \( \theta  \)\( \rightarrow 0 \),
we recover the Coulomb gauge. Similarly, one could interpolate between the axial
and the Lorentz type gauges by a gauge function such as 

\( F[A,\kappa ,\lambda ] \) = \( \frac{1}{\sqrt{\lambda }} \)\( [(1-\kappa )\partial ^{\mu }A_{\mu }+\kappa \eta .A \){]}~
~~ ~ ~~ ~~ ~~ ~~~ ~~ ~~~ ~~~ ~~ ~~ ~~ ~ ~~ ~~
~~ ~~~(1.2)

where for \( \kappa  \)=0, we recover a Lotentz-type gauge and for \( \kappa  \)=1,
we recover the axial gauge in the \( \lambda  \)\( \rightarrow 0 \) limit.
Such interpolating gauges have been employed in attempts to prove independence
of observables on the choice of the gauge.The arguments in such proofs proceed
along the same lines as those that prove the gauge independence of physical
observables under, say, a variation of the gauge parameter in the Lorentz type
gauges. 

We know that to define the Lorentz type gauges, it is necessary to define how
the poles at \( k^{2}=0 \) are treated. A correct treatment of the Lorentz
type gauges is given by adding to the Minkowski-space action a proper \( \epsilon  \)-
term \( -i\varepsilon \int d^{4}x[\frac{1}{2}A^{2}-\overline{c}c] \) ( see
section 2 for elaboration). The presence of some such a term is indispensable
in other gauges {[}even though it may not completely solve the question of their
definition{]}. It appears that the use of interpolating gauges for proving the
gauge independence has often been made without attention to a proper \( \epsilon  \)-term.
In this work, we would like to draw attention to and demonstrate several things.
We would like, first of all, to draw attention to the fact that it is very important
to deal with the \( \epsilon  \)-term carefully before interpolating gauges
can be of use in the first place. We further show that if we make a naive use
of the epsilon term \( -i\varepsilon \int d^{4}x[\frac{1}{2}A^{2}-\overline{c}c] \)
(that is appropriate for the Feynman gauge) for interpolating gauges with \emph{arbitrary
parameter values} {[}i.e. \( \alpha  \){]} , we cannot preserve gauge independence
generally {[}except in the infinitesimal neighborhood of the Feynman gauge\footnote{%
A small modification in the \( \epsilon  \)-term will allow us to change this
to {}``infinitesimal neighborhood of a Lorentz-type gauges{}''. See section
3 for more details.
} {]}. We show that preserving gauge-independence requires that {[}in general{]}
we vary the \( \epsilon  \)-term with \( \alpha  \)'s;  and that it is (probably
only) in the context of the family of Lorentz-type gauges that a simple form
can be written for all gauge parameters \( \lambda  \)'s that does not essentially
alter the treatment of poles at \( k^{2}=0 \). We do this in section 3. The
situation with a naive use of \( -i\varepsilon \int d^{4}x[\frac{1}{2}A^{2}-\overline{c}c] \)
is actually worse: In section 4, we show with an explicit example that for such
a naive use of an \( \epsilon  \)-term, we develop serious pathological behavior
in the path-integral as \( \alpha  \) is/are varied. In section 5, we work
out a simple example based on QED, where we consider if there is no other modification
of the \( \epsilon  \)-term that will remove this problem. We show that there
is no such escape available. The entire discussion in sections 3, 4 and 5 is
self-contained.We finally point out that correct way to fix the \( \epsilon  \)-term
in a path-integral in a non-Lorentz gauge is by connecting the path-integral
to the Lotentz-gauge path-integral \emph{with correct \( \epsilon  \)-}term
as has been done using the finite field-dependent BRS transformations in recent
years {[}8,1,9,10{]}. In section 6 , we summarize our conclusions.

We believe that the neglect of this issue may be the reason for the uncertainty
in the field of noncovariant gauges lasting over a long time {[}11{]}. The line
of approach suggested along the lines of references {[}8,1,9,10{]} does not
however evade this issue and has been applied successfully to axial-type gauges
{[}12{]}, planar gauges {[}13{]} and recently to the Coulomb gauge{[}14{]}.

\section{PRELIMINARIES}

\paragraph{\textmd{In this work, we propose to discuss in the path integral framework,
the interpolating gauges that interpolate between pairs of gauges. This is usually
done by considering gauge functionals that depend on one or more parameters.
We shall therefore consider the Faddeev-Popov effective action {[}FPEA{]} with
a local gauge function} \protect\( F[A(x);\alpha ]\protect \) \textmd{which
may depend on several parameters, collectively denoted by \protect\( \alpha \protect \)
.We denote this FPEA by \protect\( S_{eff}[A,c,\overline{c};\alpha ]\protect \)
which is given by }}

\( S_{eff}[A,c,\overline{c};\alpha ] \) = \( S_{0}[A] \) +\( S_{gf}[A;\alpha ] \)+\( S_{gh}[A,c,\overline{c};\alpha ] \)~
~ ~ ~~ ~~ ~ ~~ ~~ ~~ ~~~ ~~~ ~~ ~~~(2.1)

with

\( S_{gf}[A;\alpha ]= \) \( - \)\( \frac{1}{2}\int  \) d\( ^{4}x \) \( F[A,\alpha ]^{2} \)~
~~ ~~ ~~ ~ ~~ ~~ ~~ ~~~ ~~ ~~ ~~ ~~ ~~ ~~ ~~ ~~
~~ ~~ ~~~ ~~ ~~ ~~~~~~~~~~~~(2.2a)

and\footnote{%
The ghost action is always arbitrary upto a constant and, in particular, an
overall sign. The following is a convention we make.
}

\emph{S\( _{gh} \)=} \( - \)\( \int  \)d\( ^{4}x \)d\( ^{4}y \) \( \overline{c} \)\( ^{\alpha }(x) \)\( M^{\alpha \beta }[x,y;A;\alpha ]c^{\beta }(y) \)
~ ~~ ~~ ~~ ~ ~~ ~~ ~~ ~~~ ~~ ~ ~ ~~ ~~~ ~~ ~ ~~
~~ ~~ ~~ ~~ ~~ ~~(2.2b)

with

\( \int  \)d\( ^{4}y \)\( M^{\alpha \beta }[x,y;A;\alpha ]c^{\beta }(y) \)
=\( \int d^{4}y\quad \frac{\delta F^{\alpha }[A(x);\alpha ]}{\delta A^{\gamma }_{\mu }(y)} \)
D\( ^{\gamma \beta } \)\( _{\mu } \) {[}A(y){]} c\( ^{\beta }(y) \)~ ~~
~~ ~~ ~~ ~~~ ~~ ~~ ~ ~~ ~~ ~~ ~~ ~~ ~(2.3)

and

D\( ^{\alpha \beta } \)\( _{\mu } \) {[}A{]}=\( \delta ^{\alpha \beta }\partial _{\mu }+g\, f^{\alpha \beta \gamma }A_{\mu }^{\gamma } \)
~ ~~ ~~ ~~ ~ ~~ ~~ ~~ ~~~ ~~ ~~ ~~~~ ~~ ~~ ~~
~ ~~ ~~ ~~ ~~ ~~(2.3a)

Here,  \( f^{\alpha \beta \gamma } \) are the antisymmetric structure constants
of a semi-simple gauge group. 

\( S_{eff}[A,c,\overline{c};\alpha ] \) is invariant under the BRS transformations
:

\( \delta A^{a}_{\mu }(x)= \) D\( ^{\alpha \beta } \)\( _{\mu } \) {[}A(x){]}c\( ^{\beta } \)(x)
\( \delta  \)\( \Lambda  \)

\( \delta c^{\alpha }(x)= \)\( - \)\( \frac{1}{2} \) g f\( ^{\alpha \beta \gamma } \)
c\( ^{\beta } \)(x) c\( ^{\gamma } \)(x)\( \delta  \)\( \Lambda  \)

\( \delta \overline{c}^{\alpha }(x)= \) F {[}A(x); \( \alpha  \){]} \( \delta  \)\( \Lambda  \)
~ ~~ ~ ~~ ~~~ ~ ~~ ~~ ~~ ~~~~ ~ ~~ ~~ ~~ ~ ~~
~~ ~~ ~~~ ~~ ~~ ~~(2.4)

\section{\noun{The correct \protect\( \epsilon \protect \)-term for a given gauge }}

We start with the path-integral exhibiting the vacuum expectation value of a
gauge-invariant observable \( O[A] \) in the gauge characterized by the gauge
function \( F[A;\alpha ] \): 

\( <<O[A]>>\mid _{_{\alpha }}=\int D\phi \, O[A]\exp \{iS_{eff}[A,c,\overline{c}] \)
\} ~~~~~~~~~~~~~~~~~~~~~~~~~~~~(3.1)

where, \( \phi  \) collectively denotes all fields \( A,c,\overline{c} \).
The above does not completely specify how the left hand side is evaluated.In
each gauge choice, there is a residual gauge degree of freedom to a variable
extent, and this makes the propagator ill-defined.Thus in the axial-type gauges,
the propagator is ill-defined at values of \( k \) such that k.\( \eta  \)
= 0 .In the Coulomb gauge, the propagator behaves as \( \frac{1}{|\mathbf{k}|^{2}} \)
for the time-like components and this makes the Feynman integrals more ill-defined
compared to the Feynman gauge and needs an additional treatment {[}14,15{]}.
This problem is not a particularity of the noncovariant gauges only; the problem
exists for the Lorentz gauges also.In this set of gauges , the problem manifests
as a singularity at \( k^{2}=0 \). We know the solution for this problem in
the context of say the Feynman gauge. For the \emph{physical degrees of freedom},
we expect the causal prescription of Feynman, like any the propagator for a
physical degree of a massive particle.We thus expect that for the transverse
degrees of freedom, the singularity (pole) at \( k^{2}=0 \) is interpreted
by the causal replacement\footnote{%
This is not the only way to introduce a causal propagator , but the simplest
one. See appendix for more details.
} \( \frac{1}{k^{2}}\rightarrow  \)\( \frac{1}{k^{2}+i\varepsilon } \). If
, now, we want to preserve covariance, then we have to set the Feynman gauge
propagator at \( - \)\( \frac{ig_{\mu \nu }}{k^{2}+i\varepsilon } \) . WT
identities relate Green's functions involving ghosts and gauge fields and one
of the important necessities of unitarity is the cancellation of ghost and unphysical
contributions in unitarity relations {[}16{]}. This determines the \emph{sign}
of the \( \epsilon  \)-term for the ghost poles. This combined {}``prescription{}'' \footnote{%
We call this a {}`` prescription{}'' because of some arbitrariness still left
in it.
}for the pole at \( k^{2}=0 \), is indeed essentially realized by adding a term\footnote{%
Even here, only the \emph{sign} of the ghost-dependent term below is fixed.
} 

\( -i \)\( \epsilon  \)R \( \equiv  \)\{\( -i\epsilon  \)\( \int  \)d\( ^{4}x \)
\( (\frac{1}{2}A^{2}-\overline{c}c) \)\} ~~~~~~~~~~~~~~~~~~~~~~~~~~~~~~~(3.2)

to the action. In the context of general Lorentz gauges {[}\( \lambda  \) \( \neq  \)0{]},
this term gives a covariant way of handling this singularity in such a manner
that the physical degrees of freedom obey causal replacement\( \frac{1}{k^{2}}\rightarrow  \)\( \frac{1}{k^{2}+i\varepsilon } \).
This term , in particular, also implies causal treatment for \emph{all} poles
if \emph{\( \lambda  \) > 0},  since then the propagator is 

\( - \)\( \frac{i\{g_{\mu \nu }-\frac{k_{\mu }k_{\nu }}{k^{2}+i\varepsilon \lambda }\}}{k^{2}+i\varepsilon } \).
~~~~~~~~~~~~~~~~~~~~~~~~~~~~~~~~~~~~~~~~~~~~~~~~(3.3)

although covariance does not demand this.

We thus see that, in the context of the Lorentz-type gauges, there is a rationale
that leads to this \( \epsilon  \)-term. Moreover, the calculated results in
the Lorentz gauges do depend on this choice and thus it is not an arbitrary
matter to choose this term. An alteration in this {}``prescription{}'' could
change the answers for the Green's functions by O{[}1{]} and not by O(\( \epsilon  \))\footnote{%
This is not surprising since here we are altering the value of the integrands
in a region where a large contribution comes.
}! For example, changing the sign of the \( \epsilon  \)-term, \emph{which may
naively be thought as a \( O(\varepsilon ) \) change in the action}, changes
the Green's functions, even for physical degrees of freedom, from causal to
anti-causal ( i.e. from physically meaningful to unphysical)! Then,  it is far
from obvious that some other \( O(\varepsilon ) \) change in this term cannot
drastically alter the Green's functions and in particular wrench gauge-independence.
(What, however, may not be obvious is that such a \( O(\varepsilon ) \) change
can in fact arise from parameter variations: that we will work out soon). Thus,
it is evident that a correct treatment of this \( \epsilon  \)-term is desired\footnote{%
In other words, we are emphasizing that the proper \( \epsilon  \)-term is
not a trifle O(\( \epsilon  \)) matter but an matter deserving \( O(1) \)
attention!
} in any given gauge if gauge-independence is to be preserved. It is by no means
obvious that such a term as (3.2) will do the job correctly for any other class
of gauges, and yet preserve gauge independence of observables. We wish to explicitly
demonstrate , that this is in fact not the case: the term (3.2) is an acceptable
\( O(\varepsilon ) \)-term for the Lorentz-type gauges but generally not for
other class of gauges. We shall thereafter make comments, in Sec. 4, on the
correct way out of this situation that has already been proposed in {[}8,1{]}
and followed up in {[}9,10{]}. {[}We note that in discussing the gauge-independence
of S-matrix elements, it is necessary to consider the \emph{on-shell} Green's
functions which are directly accessible only in Minkowski space ; and this makes
us necessary to deal with the \( \epsilon  \)-term{]}. 

To begin with, let us consider a gauge function \( F[A,\alpha ] \) that connects
to the Feynman gauge. Let us suppose that for certain value(s) of the parameter(s)
\( \alpha  \)=\( \alpha  \)\( _{0} \), the gauge-fixing term corresponds
to the Feynman gauge.Then, 

\( <<O[A]>>\mid _{_{\alpha _{0}}}=\int D\phi \, \, O[A]\exp \{iS_{0}[A]-\frac{i}{2}\int  \)
d\( ^{4}x \) \( F[A(x),\alpha _{0}]^{2} \)+\( iS_{gh}[A,c,\overline{c} \);
\( \alpha  \)\( _{0} \){]}

~~~~~~~~~~~~~~~~~~~~~~~~~~~~~~~~~~~~~~+
\( \int  \)d\( ^{4}x \) \( \epsilon  \) (\( \frac{1}{2}A^{2}-\overline{c} \)c)
\}~~~~~~~~~~~~~~~~~(3.4)

correctly describes the value of the vacuum expectation value of O{[}A{]} in
the Feynman gauge. We expect the Feynman rules in a gauge described by parameter(s)
\( \alpha  \)-\( \delta  \)\( \alpha  \) are such that the vacuum expectation
value of O{[}A{]} is unchanged.We would like to know what should happen, if
at all, to the \( \epsilon  \)-term as \( \alpha  \) is varied from \( \alpha  \)\( _{0} \)
\( \rightarrow  \) \( \alpha  \)\( _{0} \)-\( \delta  \)\( \alpha  \) so
that this remains valid. To do this , we perform the field transformation below:

\( A'^{\alpha }_{\mu }(x)-A^{\alpha }_{\mu }(x)\equiv  \)\( \delta A^{\alpha }_{\mu }(x)= \)
\( i \)D\( ^{\alpha \beta } \)\( _{\mu } \) {[}A(x){]}c\( ^{\beta } \)(x)
\( \int  \)\( d^{4}z \) \( \overline{c}^{\gamma } \)\( (z) \)\( \frac{\partial F^{\gamma }[A(z);\alpha ]}{\partial \alpha }\mid _{_{a_{0}}} \)\( \delta  \)\( \alpha  \)

\( \delta c^{\alpha }(x)=- \)\( i \)\( \frac{1}{2} \) g f\( ^{\alpha \beta \delta } \)
c\( ^{\beta } \)(x) c\( ^{\delta } \)(x) \( \int  \)\( d^{4}z \) \( \overline{c}^{\gamma } \)\( (z) \)\( \frac{\partial F^{\gamma }[A(z);\alpha ]}{\partial \alpha }\mid _{_{a_{0}}} \)\( \delta  \)\( \alpha  \)

\( \delta \overline{c}^{\alpha }(x)= \)\( i \)F\( ^{\alpha } \){[}A(x); \( \alpha _{0} \){]}\( \int  \)\( d^{4}z \)\( \overline{c}^{\gamma } \)\( (z) \)\( \frac{\partial F^{\gamma }[A(z);\alpha ]}{\partial \alpha }\mid _{_{a_{0}}} \)\( \delta  \)\( \alpha  \)
~~~~~~~~~~~~~~~~~~~~~~~~~~~~~~~~~(3.5)

We recognize the above simply as a BRS transformation\footnote{%
We have called this infinitesimal field-dependent BRS transformation or an IFBRS
for short {[} 9{]}.
} with a \emph{field-dependent} \( \delta  \)\( \Lambda  \) = \( i \)\( \int  \)\( d^{4}z \)
\( \overline{c} \)\( (z) \)\( \frac{\partial F[A(z);\alpha ]}{\partial \alpha }\mid _{_{a_{0}}} \)\( \delta  \)\( \alpha  \)
. The action S\( _{eff}[A,c,\overline{c}] \) is in fact invariant under this
{[}8{]}.But, the transformation is \emph{nonlocal} and leads to a nontrivial
Jacobian.The transformation being infinitesimal, only the diagonal terms of
the transformation matrix matter, and on account of antisymmetry of the structure
constants,  only the field dependences of \( \delta  \)\( \Lambda  \) = \( i \)\( \int  \)\( d^{4}z \)
\( \overline{c} \)\( \frac{\partial F}{\partial \alpha }\mid _{_{a_{0}}} \)\( \delta  \)\( \alpha  \)
contributes.

Defining the Jacobian for the infinitesimal transformation (3.5) as 

D\( \phi  \) \( \equiv  \) D\( \phi ' \) J\( ^{-1} \) \( \equiv  \) D\( \phi ' \){[}1\( - \)\( \Delta  \)J{]}~~~~~~~~~~~~~~~~~~~~~~~~~~~~~~~~~~~~~~~~~~~~~~~~~(3.6)

\( \Delta  \)J is given by 

\( -\Delta  \)J = \( -\int  \)\( d^{4}x \) \{ \( \sum  \)\( _{_{_{\alpha ,\mu }}} \)
\( \frac{\delta A^{'\alpha }_{\mu }(x)}{\delta A^{\alpha }_{\mu }(y)} \)\( \mid _{_{_{x=y}}} \)
\( - \)\( \sum  \)\( _{_{_{\alpha }}} \) \( \frac{\delta \overline{c}'^{\alpha }(x)}{\delta \overline{c}^{\alpha }(y)} \)\( \mid _{_{_{x=y}}} \)\}

= \( i \)\( \int  \)\( d^{4}x \) \( \int  \)\( d^{4}z \) \( \overline{c} \)\( ^{\gamma }(z) \)\( \frac{\partial }{\partial \alpha } \)\{\( \frac{\delta F^{\gamma }[A(z);\alpha ]}{\delta A^{\alpha }_{\mu }(y)} \)\}\( \mid _{_{a_{0}}} \)D\( ^{\alpha \beta } \)\( _{\mu } \)
{[}A(y){]}\( c^{\beta }(y) \)\( \delta  \)\( \alpha  \) \( +i\int d^{4}xF^{\gamma }(\alpha _{0})\frac{\partial F^{\gamma }}{\partial \alpha } \)\( \mid _{_{_{\alpha _{0}}}}\delta \alpha  \)~~~~~~~~~~~~~~~~~~~~(3.7)

Thus, 

\( <<O[A]>>\mid _{_{\alpha _{0}}}=\int D\phi 'O[A']\exp \{ \)\( iS_{eff}[A',c',\overline{c}' \);
\( \alpha  \)\( _{0} \){]}+ \( \int  \)d\( ^{4}x \) \( \epsilon  \) R'(A',c',\( \overline{c} \)'
) \( - \) \( \Delta  \)J\} \( +O[(\delta \alpha )^{2}] \)~~~~~~~~~~~~~~~~(3.8)

In (3.8) above, we note two things: 

(a) We note that \( -\Delta  \)J simply changes the form of the gauge-fixing
and the ghost terms as implied by:

iS\( _{eff} \) {[} A,c,\( \overline{c} \); \( \alpha  \)\( _{0} \){]} \( - \)\( \Delta  \)J
\( \equiv  \)\( \,  \)iS\( _{eff} \) {[} A,c,\( \overline{c} \); \( \alpha  \)\( _{0} \)-\( \delta  \)\( \alpha  \){]}~~~~~~~~~~~~~~~~~~~~~~~~~~~~~~~~~~(3.9)

which is readily verified. 

(b) Further, we note that \emph{there is a change in the \( \epsilon  \)-terms}
also:

\( \epsilon  \)R(A,c,\( \overline{c} \)) \( \equiv  \)\( \epsilon  \)R'(A',c',\( \overline{c} \)'
) \( \equiv  \)\( \epsilon  \){[}R(A',c',\( \overline{c} \)' ) + \( \delta  \)R(A',c',\( \overline{c} \)'
){]} ~~~~~~~~~~~~(3.10a)

~~~~~~~~~~~~~~~~~~= \( \epsilon  \)\( \int  \)\( d^{4}x \)
( \( \frac{1}{2}A'^{2}-\overline{c}' \)c') 

~~~~~~~~~~~~~~~\( - \) \( \epsilon  \)\( \int  \)d\( ^{4}x \)
\{\( iA_{\mu } \)D\( ^{\mu }c \) \( \int  \)d\( ^{4}z \) \( \overline{c} \)\( \frac{\partial F}{\partial \alpha } \)\( \mid _{_{_{\alpha _{0}}}}\delta \alpha  \)
\( -i \)\( F(\alpha _{0}) \)\( \int  \)d\( ^{4}z \)\( \overline{c} \)\( \frac{\partial F}{\partial \alpha } \)\( \mid _{_{_{\alpha _{0}}}}\delta \alpha  \)
\( c \) 

~~~~~~~~~~~~~\( + \) \( \overline{c} \)(x) \( \left( -\frac{i}{2}g\, f^{\alpha \beta \gamma }c^{\beta }c^{\gamma }\int d^{4}z\overline{c}(z)\frac{\partial F}{\partial \alpha }\mid _{_{\alpha _{0}}}\delta \alpha \right)  \)\}~~~~~~~~~~~~~~~~~~~~~~~~~~~~(3.10b)

~~~~~~~~~~~= \( \epsilon  \)\( \int  \)d\( ^{4}x \) \{ (\( \frac{1}{2}A'^{2}-\overline{c}' \)c')\( +i \)
\( \int  \)d\( ^{4}z \) \( \overline{c} \)\( \frac{\partial F}{\partial \alpha } \)\( \mid _{_{_{\alpha _{0}}}}\delta \alpha  \)\( A_{\mu } \)\( \partial  \)\( ^{\mu }c \)
+ \( i \) \( \int  \)d\( ^{4}z \)\( \overline{c} \)\( \frac{\partial F}{\partial \alpha } \)\( \mid _{_{_{\alpha _{0}}}}\delta \alpha  \)
\( F(\alpha _{0}) \)\( c \)\} 

~~~~~~~~~~~+ \( O[g] \)~~~~~~~~~~~~~~~~~~~~~~~~~~~~~~~~~~~~~~~~~~~~~~~~~~~~~~~~~~~~~~(3.10c)

We note that in the special case, F(\( \alpha  \)\( _{0} \)) = \( \partial .A \)
{[} the Feynman gauge{]} the last two terms combine to have an integral over
a 4-divergence: \( \int d^{4}x \) \( \partial  \)\( ^{\mu }(A_{\mu }c) \).
{[}This arises from the fact that R is invariant under the abelian part of BRS
transformations of the \emph{Feynman gauge} {]}.Thus, when a slight variation
is made around the Feynman gauge , the quadratic part of \( \epsilon  \)-term
does not alter to the first order in \( \delta  \)\( \alpha  \). This, in
fact, can be shown to be true for the entire class of the Lorentz gauges with
a gauge parameter \( \lambda  \) > 0 with a suitable modification of \( \epsilon  \)R
of (3.2), which is not completely determined. All we need to recall is that
the causal prescription does not depend on {}``\( \epsilon  \){}'' but on
its sign and the fact that \( \epsilon  \) \( \rightarrow  \) 0\( ^{+} \).We
can as well choose the quadratic \( \epsilon  \)-term, to be added to the action,
as 

\( -i\int  \)d\( ^{4}x \) \( \epsilon  \) (\( \frac{1}{2}A^{2}-\sqrt{\lambda }\overline{c} \)c)
~~~~~~~~~~~~~~~~~~~~~~~~~~~~~~~~~~~~~~~~~~~~~~(3.11)

and this will as well provide the correct causal prescription above for \( \lambda >0 \).Thus
with this modification, the additional terms , quadratic in A become {[}we note
that here F{[}\( A;\alpha  \){]} \( \equiv  \)F{[}\( A;\lambda  \){]} = \( \frac{\partial .A}{\sqrt{\lambda }} \){]},

=\( - \) \( \epsilon  \) \{ \( -i \) \( \int  \)d\( ^{4}z \) \( \overline{c} \)\( \frac{\partial F}{\partial \alpha } \)\( \mid _{_{_{\alpha _{0}}}}\delta \alpha  \)
\( \int  \)d\( ^{4}x \) {[}\( A_{\mu } \)\( \partial  \)\( ^{\mu }c \)
+\( \sqrt{\lambda } \) \( \frac{\partial .A}{\sqrt{\lambda }} \)c{]} \( \equiv 0 \).~~~~~~~~~~~~~~~~~~~~~~(3.12)

Equivalently, (3.11) is invariant under the abelian part of BRS for \emph{general
Lorentz gauges.} It is on account of this property of the Lorentz gauges that
we do not need to adjust the \( \epsilon  \)-term (3.11) as we vary the gauge
parameter \( \lambda  \) is varied between 0 and infinity\footnote{%
We do not necessarily require positive \( \lambda  \) here: we could instead
take the ghost term as \( - \)\( \sqrt{\lambda } \)\( \overline{c}Mc \).
This will cancel \( \frac{1}{\sqrt{\lambda }} \)in \( M \). We also replace
R by \( \int d^{4}x(\frac{1}{2}A^{2}-\lambda \overline{c}c) \).Then the above
invariance is valid for all \( \lambda  \) .
}. The causal prescription is also not altered by the change in \( \epsilon  \)-term
that arises as \( \lambda  \) is varied.

We would now like to discuss the possibility that an expression of the type
of (3.4) \emph{with the same \( \epsilon  \)-term} gives correctly \( <<O[A]>> \)
when \( F[A,\alpha ] \) refers to some gauge other than the Lorentz-type gauge.
This value \( <<O[A]>> \) must then be independent of \( \alpha  \)'s and
must, in particular, agree with the Lorentz gauge result.

To give examples, we could consider,

1) Interpolating gauge used by Doust to interpolate between the Feynman and
the Coulomb gauge.It has 

\( F[A,\theta ] \) = \( [\theta \partial ^{0}A_{0}-\frac{1}{\theta } \)\( \partial  \)\( _{i}A_{i}] \)~~~~~~~~~~~~~~~~~~~~~~~~~~~~~~~~~~~~~~~~~~~~~~~(3.13)

2) Interpolating gauges between the axial and the Lorentz-type gauges:

\( F[A,\kappa ] \) = \( \frac{1}{\sqrt{\lambda }} \)\( [(1-\kappa )\partial ^{\mu }A_{\mu }+\kappa \eta .A] \)
\( \qquad \qquad  \) 0 \( \leq  \)\( \kappa  \)\( \leq  \)1~~~~~~~~~~~~~~~~~~~(3.14)

We then see that these extra terms ( to \( O[g^{0}] \)) now add up to 

\( - \)\( \epsilon  \) \{ \( i \) \( \int  \)d\( ^{4}z \) \( \overline{c} \)\( \frac{\partial F}{\partial \alpha } \)\( \mid _{_{_{\alpha _{0}}}}\delta \alpha  \)
\( \int  \)d\( ^{4}x \) {[}\( \partial  \).\( A \)c -F{[}A,\( \alpha  \){]}c{]}\}~~~~~~~~~~~~~~~~~~~~~~~~(3.15)

and this would not vanish in general unless F{[}A,\( \alpha  \){]} refers to
the Feynman gauge. We shall see in the next section, the explicit effects of
this term on the free gauge boson propagator and that we cannot ignore this
term. How this works is not so easily anticipated, and is best understood after
some calculations in the next section are first performed.

\section{\noun{concrete evaluation of the effects of the \protect\( \epsilon \protect \)-term}}

In this section, we wish to explore the possibility that a path-integral of
the form 

W{[}J; \( \alpha  \){]} = \( \int  \)D\( \phi  \) \( \exp  \)\{i S\( _{eff} \){[}A,c,\( \overline{c} \);
\( \alpha  \){]} + \( \epsilon R \)+ i\( \int J^{\mu }A_{\mu } \)d\( ^{4}x \)\}~~
~~ ~~ ~~~~~ ~~~~ ~~ ~~ ~~ ~~ (4.1)

with \emph{the same \( \epsilon  \)}-term as in (3.2) correctly generates the
Green's functions for the gauge \( F[A;\alpha ] \) \emph{other than the Lorentz
family of gauges.} So, let us assume that this is so to begin with.We next consider
the generating functional 

W{[}J; \( \alpha  \)- \( \delta \alpha  \){]} \( \equiv  \) \( \int  \)D\( \phi  \)
\( \exp  \)\{i S\( _{eff} \){[}A,c,\( \overline{c} \); \( \alpha  \)\( -\delta  \)\( \alpha  \){]}
+ \( \epsilon  \){[}R+ \( \delta  \)R{]}+ i\( \int J^{\mu }A_{\mu } \)d\( ^{4}x \)\}~~
~~ ~~~ ~~ ~~ (4.2)

Here, we are taking the effective action associated with the parameter(s) \( \alpha  \)\( -\delta  \)\( \alpha  \)
and the \( \epsilon  \)-term \( \epsilon  \){[}R+ \( \delta  \)R{]} obtained
in the last section.We had noted, in section 3, that to preserve the gauge-independence
\emph{}of the vacuum expectation value of an observable, this term was required\emph{.
Suppose we assume} that at \( \delta \alpha  \)=0, the generating functional
in fact gives the correct Green's functions; i.e., the term \( \epsilon  \)R
is, in fact, the right \( \epsilon  \)-term, then according to the discussion
of the last section, (4.2) should generate the correct Green's functions for
the parameters \( \alpha  \)\( -\delta  \)\( \alpha  \). We now have to note
the presence of the extra term \( \epsilon  \) \( \delta  \)R in (4.1).We
would like to know what effect this term has on the Green's functions in the
gauge \( \alpha  \)\( -\delta  \)\( \alpha  \) . In the special case of the
Lorentz type gauges as the gauge parameter \( \lambda  \) was varied, we saw
that the effect, as \( \epsilon  \)\( \rightarrow  \) 0\( ^{+} \), was none;
 we shall however like to show that that is rather an exception than a rule.

We shall first explore the effect of this term on the \emph{tree} propagator
for a gauge field A. We note that the factor \( \exp  \)\{\( \epsilon  \)
\( \delta  \)R\} inside the path-integral can be written, to first order in
\( \delta  \)\( \alpha  \), as

W{[}J{]} = \( \int  \)D\( \phi  \) \( \left( 1+\varepsilon \delta R+O[\left( \delta \alpha \right) ^{2}]\right)  \)\( \exp  \)\{i
S\( _{eff} \){[}A,c,\( \overline{c} \); \( \alpha  \)\( +\delta  \)\( \alpha  \){]}
+ \( \epsilon  \)R+ i\( \int J^{\mu }A_{\mu } \)d\( ^{4}x \)\}~~ ~~ ~~
~~ (4.2a)

We recall that 

\( \delta  \)R = \( - \)\{ \( i \) \( \int  \)d\( ^{4}z \) \( \overline{c} \)\( \frac{\partial F}{\partial \alpha } \)\( \mid _{_{_{\alpha _{0}}}}\delta \alpha  \)
\( \int  \)d\( ^{4}x \) \{\( \partial  \).\( A \)c -F{[}A,\( \alpha  \){]}c\}\( +O[g] \)~
~~ ~~ ~~~~ ~~ ~~ ~~ ~~~~ ~~ ~~ (4.3)

To evaluate the effect of this term on the bare propagator, it is possible to
evaluate this term by first performing the ghost integration by making use of
:

\( \int DcD\overline{c} \) \( c^{\alpha }(x)\overline{c}^{\beta }(z) \) \( \exp  \)
\{ \( i \) \( S_{gh}[A,c,\overline{c};\alpha ]-\varepsilon \int \overline{c}cd^{4}x\} \)

= \( -iM^{'-1\alpha \beta }(x,z;A;\alpha ) \) \( \int DcD\overline{c} \) \( \exp  \)
\{ \( i \) \( S_{gh}[A,c,\overline{c};\alpha ]-\varepsilon \int \overline{c}cd^{4}x\} \)~~
~~ ~~~~ ~~~~~ ~~~~ ~~ (4.4)

with \( M'=M-i\varepsilon  \), and M is given by (2.3).Thus,

W{[}J{]} = \( \int  \)D \( \phi  \) \( \left( 1-i\varepsilon \int d^{4}z\frac{\partial F^{\beta }}{\partial \alpha }\mid _{_{\alpha _{0}}}\delta \alpha (\partial .A-F)^{\alpha }(x)i\int d^{4}xM^{'-1\alpha \beta }(x,z;A;\alpha _{0})+O[\left( \delta \alpha \right) ^{2}]+O[g]\right)  \)
\( \exp  \)\{i S\( _{eff} \){[}A,c,\( \overline{c} \); \( \alpha  \)\( +\delta  \)\( \alpha  \){]}
+ \( \epsilon  \)R+ i\( \int J^{\mu }A_{\mu } \)d\( ^{4}x \)\}~~ ~~ ~~
~~~~ ~~ ~~~ ~ ~~~ ~~~~ ~~~ ~~~~ ~~ ~~~ ~~ ~~
(4.5)

We now exponentiate the result so obtained to find:

W{[}J{]} = \( \int  \)D\( \phi  \)\( \exp  \)\{iS\( _{eff} \){[}A,c,\( \overline{c} \);
\( \alpha  \)\( -\delta  \)\( \alpha  \){]}+\( \epsilon  \)R+i\( \int J^{\mu }A_{\mu } \)d\( ^{4}x \)

~ ~~ ~~ ~~~ ~~ ~~~~ ~~+ \( \left( \varepsilon \int d^{4}z\frac{\partial F^{\beta }}{\partial \alpha }\mid _{_{\alpha _{0}}}\delta \alpha \int d^{4}x(\partial .A-F)^{\alpha }(x)M^{'-1\alpha \beta }(x,z;A;\alpha _{0})\right)  \)\}~
~~ ~~ ~~ (4.6)

Now the propagator in the gauge with parameters \( \alpha  \)-\( \delta  \)\( \alpha  \)
is determined by the quadratic part of the net action including the \( \epsilon  \)-terms:

S\( ^{(2)}_{0}[A] \)+S\( _{gf}-i\varepsilon \frac{1}{2}\int A^{2}d^{4}x \)
- \( \left( i\varepsilon \int d^{4}z\frac{\partial F^{\beta }}{\partial \alpha }\mid _{_{\alpha _{0}}}\delta \alpha (\partial .A-F)^{\alpha }(x)\int d^{4}xM'^{-1\alpha \beta }_{0}(x,z;\alpha _{0})\right)  \)

~ ~~ ~~ ~~~\( + \) O {[}(\( \delta  \)\( \alpha  \))\( ^{2} \){]}.~
~~ ~~ ~~ ~ ~~ ~~ ~~~ ~~~ ~~ ~~ ~ ~~ ~~ ~~~ ~ ~~
~~ ~~ (4.7)

Here, \( M'_{0} \) refers to \( M' \) at g=0.To see the concrete effects,
we need to take a specific example.Suppose we take the interpolating gauge term
of Doust {[} we assume that \( \theta  \)\( \neq 0 \){]},

\( F[A,\theta ] \) = \( [\theta \partial ^{0}A_{0}-\frac{1}{\theta } \)\( \partial  \)\( _{i}A_{i}] \)~~
~~ ~~~~ ~~ ~~ ~~ ~~~ ~~ ~~~ ~~~ ~~~ ~~~ ~~~
~~ ~~ ~~ ~~ (4.8a)

M{[}A, \( \theta  \){]} \( \equiv  \)\( \frac{\delta F}{\delta A_{\mu }}D_{\mu } \)
= \( [\theta \partial ^{0}D_{0}-\frac{1}{\theta } \)\( \partial  \)\( _{i}D_{i}] \)~~
~~~~ ~~ ~~ ~~ ~~~ ~~ ~~~ ~~ ~~ ~~ ~~ ~~ ~~ (4.8b)

M'\( ^{\alpha \beta } \)\( _{0} \){[}\( \theta  \){]} = \( [\theta \partial ^{2}_{0}-\frac{1}{\theta } \)\( \nabla ^{2} \)\( -i\epsilon  \){]}\( \delta  \)\( ^{\alpha \beta } \)~
~~ ~~ ~~ ~~~ ~~ ~~ ~~~ ~~ ~~ ~~ ~~ ~~ ~~ ~~ (4.8c)

\( <y| \)M'\( _{0} \)\( ^{-1\alpha \beta } \){[}\( \theta  \){]}\( |x>\equiv M_{0}^{'-1\alpha \beta }(x,y;\theta ) \)
\( = \) \( \delta  \)\( ^{\alpha \beta } \)\( \frac{1}{\theta \partial ^{2}_{0}-\frac{1}{\theta }\nabla ^{2}-i\varepsilon } \)\( \delta  \)\( ^{4} \)(x-y) 

~~ ~~~ ~~~ ~=\( -\delta  \)\( ^{\alpha \beta } \)\( \frac{1}{(2\pi )^{4}}\int  \)
\( \frac{d^{4}k}{\theta k^{2}_{0}-\frac{1}{\theta }|\mathbf{k}|^{2}+i\varepsilon }\exp  \)\{\( -ik.(x-y)\} \)~
~~ ~~ ~~ ~~ ~~ ~~ ~~ (4.8d)

The extra term in (4.7 ) is then 

\( \left( -i\varepsilon \int d^{4}z\frac{\partial F^{\beta }(z)}{\partial \theta }\mid _{_{\theta _{0}}}\delta \theta \int d^{4}x(\partial .A-\theta \partial ^{0}A_{0}+\frac{1}{\theta }\partial _{i}A_{i})^{\alpha }(x)M_{0}^{'-1\alpha \beta }(x,z;\alpha _{0})\right)  \)

=\( \left( -i\varepsilon \delta \theta \int d^{4}z\{\partial ^{0}A_{0}+\frac{1}{\theta ^{2}}\partial _{i}A_{i}\}(z)\int d^{4}x\{(1-\theta )\partial ^{0}A_{0}+(\frac{1}{\theta }-1)\partial _{i}A_{i}\}^{\alpha }(x)M_{0}^{'-1\alpha \beta }(x,z;\alpha _{0})\right)  \)

~~ ~~ ~~ ~~~~ ~~ ~~ ~~~ ~~ ~~ ~~~ ~~ ~~ ~~~~
~~ ~~~~~~ ~~ ~~~ ~~~~~~ ~~~~~~~ ~~ ~~ ~~ (4.9)

This term, quadratic in A, has to be taken into account in evaluating the gauge
propagator. As an illustration of the effect of the new contribution to the
\( \epsilon  \)-term, we shall focus our attention on the propagator for the
(0,0) component. We call the momentum space quadratic form arising from \( [S_{eff}-\frac{i\varepsilon }{2}\int d^{4}x \)A\( ^{2} \)
{]} as Z\( _{\mu \nu } \) .We also call the \emph{total} quadratic form arising
from the \emph{total action} including the term (4.9) as \( Z'_{\mu \nu } \)
.We then have, 

\( Z'_{\mu \nu } \)(\( \theta  \)-\( \delta  \)\( \theta  \)) = Z\( _{\mu \nu } \)(\( \theta  \)-\( \delta  \)\( \theta  \))
+ \( \epsilon  \)\( \delta  \)\( \theta  \) Z\( ^{1}_{\mu \nu } \)(\( \theta  \))
~ ~ ~~ ~~ ~~ ~~ ~~~~ ~~ ~~ ~~ ~~ ~~ ~~ ~~ ~~ ~~
~~ ~~ (4.10)

where the last term has arisen from (4.9). Noting that Z\( _{\mu \nu } \)(\( \theta  \))
for any \( \theta  \) has a block diagonal structure {[}7{]},  \( Z'_{\mu \nu } \)(\( \theta  \)-\( \delta  \)\( \theta  \))
has the structure,

\vspace{0.5001cm}
{\centering \begin{tabular}{|c|c|}
\hline 
\( Z_{00}+\varepsilon \delta \theta Z^{1}_{00} \)&
\( \varepsilon \delta \theta Z^{1}_{0i} \)\\
\hline 
\( \varepsilon \delta \theta Z^{1}_{i0} \)&
\( Z_{ij}+\varepsilon \delta \theta Z^{1}_{ij} \)\\
\hline 
\end{tabular}\par}
\vspace{0.5001cm}

~ ~ ~~ ~~ ~~ ~~ ~~ ~~ ~~~ ~ ~~ ~~ ~~ ~~ ~~ ~ ~
~~ ~~ ~ ~ ~~ ~~ ~~ ~~ ~~ ~~ ~~~ ~ ~~ ~~ ~~ ~~
~~ ~~ ~~ ~~ ~(4.11)

and consequently, its inverse has the form 

\vspace{0.5001cm}
{\centering \begin{tabular}{|c|c|}
\hline 
\{\( Z_{00}+\varepsilon \delta \theta Z^{1}_{00}\}^{-1} \)&
\( O[\varepsilon \delta \theta ] \)\\
\hline 
\( O[\varepsilon \delta \theta ] \)&
\{\( Z+\varepsilon \delta \theta Z^{1}\} \)\( _{ij} \)\( ^{-1} \)\\
\hline 
\end{tabular}\par}
\vspace{0.5001cm}

~~ ~~ ~~ ~~~ ~ ~~ ~~ ~~ ~~ ~~ ~~ ~~ ~~ ~~ ~~ ~~
~~ ~~~ ~ ~~ ~~ ~~ ~~ ~~ ~~ ~~ ~~ ~~~ ~ ~~ ~~ ~~
~~ ~~ ~~ ~~ ~~ ~(4.12)

{[}modulo terms of \( O[(\varepsilon \delta \theta )^{2}] \). The the propagator
for the timelike component is 

G\( _{00} \) =\( i \) \{\( Z_{00}+\varepsilon \delta \theta Z^{1}_{00}\}^{-1} \)~~
~ ~~ ~~ ~~ ~~ ~~ ~~ ~~ ~~~ ~~ ~~ ~~~ ~~~ ~~ ~~
~~ ~~ ~~ ~~ ~(4.13)

We note 

\( Z_{00} \) = |\textbf{k|\( ^{2} \)\( - \)(\( \theta -\delta \theta ) \)\( ^{2} \)\( k_{0}^{2} \)\( - \)\( i\varepsilon  \)
\( \equiv  \)\( -[(\theta -\delta \theta )X+i\varepsilon ] \) \( \equiv  \)\( -[\theta 'X+i\varepsilon ] \)}
~ ~ ~~ ~~ ~~ ~~ ~~ ~~ ~~ ~~ ~(4.14)

Here,

\( \theta '\equiv \theta -\delta \theta  \) and \( X\equiv \theta 'k^{2}_{0}-\frac{1}{\theta '}|\mathbf{k}|^{2} \)~
~ ~~ ~~ ~~ ~~ ~~ ~~ ~~ ~~ ~~ ~~ ~~ ~~ ~~ ~~ ~(4.14a)

\( \varepsilon \delta \theta Z^{1}_{00} \)\textbf{\( \equiv  \)}2\( i\epsilon  \)\( \delta  \)\( \theta  \)(1-\( \theta ' \))\( \frac{k_{0}^{2}}{X+i\varepsilon } \)
+ \( O[\varepsilon (\delta \theta )^{2}] \) ~ ~~ ~~ ~~ ~~ ~~ ~~~~
~~ ~~ ~~ ~~ ~(4.15)

and thus the propagator for the timelike component reads,

G\( _{00} \) =\( i \) \{\( Z_{00}+\varepsilon \delta \theta Z^{1}_{00}\}^{-1} \)=\( \frac{-i}{\theta 'X+i\varepsilon \left( 1-2\delta \theta [1-\theta ']\frac{k_{0}^{2}}{X+i\varepsilon }\right) } \)
+ \( O[\varepsilon (\delta \theta )^{2}] \) ~~ ~~ ~~ ~~ ~~ ~(4.16)

We note the net \( \epsilon  \)-terms in \( Z_{00}+\varepsilon \delta \theta Z^{1}_{00} \)
, viz. 

\( i\varepsilon  \) \( \left( 1-2\delta \theta (1-\theta ')\frac{k_{0}^{2}}{X+i\varepsilon }\right)  \)~~
~~ ~~ ~~ ~~ ~~~ ~~ ~~ ~~ ~~ ~~ ~~~ ~~ ~~ ~~ ~~
~~~~ ~~ ~~ ~(4.17)

has now received an extra contribution, which though proportional to \( \delta  \)\( \theta  \),
\emph{is very sensitive in magnitude near a k\( _{0} \) corresponding to X=0
provided \( \theta  \) \( \neq  \)1{[}i.e. we are not at the Feynman gauge{]}.}
For example, for \( X=\varepsilon  \), the net \( \epsilon  \)-term is 

\( i\varepsilon -\delta \theta (1-\theta ')k^{2}_{0}(1-i) \)

and its imaginary part has a term that cannot be made arbitrarily small as \( \epsilon  \)\( \rightarrow  \)0
and can overwhelm the original term. This is the source of trouble! The \( \epsilon  \)-term
now is dependent on \emph{two unrelated} small quantities \( \epsilon  \) and
\( \delta  \)\( \theta  \).

To explore the trouble further, we express\footnote{%
In the following, we drop the prime on \( \theta  \) for convenience.
}

Y\textbf{\( \equiv  \)}\( \theta  \)X +\( i\varepsilon  \) \( \left( 1-2\delta \theta (1-\theta )\frac{k_{0}^{2}}{X+i\varepsilon }\right)  \)
\textbf{\( \equiv  \)\( ReY+iImY \)} ~ ~~ ~~ ~~ ~~ ~~ ~~ ~~~
~~ ~~ ~~ ~~ ~(4.18)

We have to solve for the roots of the denominator , viz. 

Y\textbf{\( \equiv  \)}\( \theta  \)X +\( i\varepsilon  \) \( \left( 1-2\delta \theta (1-\theta )\frac{k_{0}^{2}}{X+i\varepsilon }\right)  \)
\( =0 \).~~ ~~ ~~ ~~ ~~ ~~ ~~ ~~ ~~ ~~~ ~~ ~~ ~~
~~ ~~ ~(4.19)

Multiplying by \( (X+i\varepsilon ) \), and substituting for \( k_{0}^{2} \)
in terms of \( X \) and \textbf{|k| we} find ,

\( \theta X^{2}+aX+b=0, \)~ ~~ ~~ ~~ ~~ ~~ ~~~ ~~ ~~ ~~
~~~ ~~ ~~~~ ~~ ~~ ~~~~ ~~~~ ~~ ~~ ~~ ~~ ~(4.20)

with 

\( a=i\varepsilon (1+\theta -\frac{2\delta \theta (1-\theta )}{\theta }) \)
and \( b=-\{\varepsilon ^{2}+i\varepsilon \frac{2\delta \theta (1-\theta )}{\theta ^{2}}|\mathbf{k}|^{2}\} \)~
~~ ~~ ~~ ~~~~ ~~ ~~ ~~ ~ ~(4.21)

The roots of the above quadratic are 

\( X=\frac{-a\pm \sqrt{a^{2}-4b\theta }}{2\theta } \)\( \equiv X_{1,2} \)~
~~~ ~~ ~~ ~~~ ~~ ~~~ ~~~~ ~~ ~~~ ~~~~ ~~~~ ~~
~~~ ~~ ~~ ~(4.22)

We note that the discriminant \( a^{2}-4b\theta  \) contains terms that are
of \( O(\varepsilon ^{2}) \) as well as \( O(\varepsilon \delta \theta ) \)
and the relative magnitudes of these terms is variable as \( \epsilon  \) is
varied for a fixed \( \delta \theta  \). We note in particular that for \( \epsilon  \)
sufficiently small {[} which is the limit we are particularly interested and
which we will denote for brevity by {}``\( \epsilon  \) <\textcompwordmark{}<
\( \delta  \)\( \theta  \){}'' {]}, it is the \( O(\varepsilon \delta \theta ) \)
term that dominates and thus, we have, for |\textbf{k| >} 0 and \( \theta \neq 1 \),

\( X_{1}\simeq -X_{2}\simeq \sqrt{2i\varepsilon \delta \theta (1-\theta )}\frac{|\mathbf{k}|}{\sqrt{\theta }} \)\( \equiv  \)\( \left( 1+i\right) \frac{2|\mathbf{k}|}{\sqrt{\theta }}\sqrt{\varepsilon \delta \theta (1-\theta )} \)
~~ ~~ ~~~ ~~~~ ~~~~ ~~ ~(4.23)

We note that in this case,  |X\( _{1,2} \)|>\textcompwordmark{}>\( \epsilon  \).

On the other hand at the other extreme limit when \( \epsilon  \) is sufficiently
large compared to \( \delta  \)\( \theta  \) {[}which we denote for brevity
by \( \epsilon  \)>\textcompwordmark{}> \( \delta \theta  \){]} the term \( O(\varepsilon \delta \theta ) \)
is negligible and we find

\( X_{1}\sim -i\varepsilon /\theta  \) and \( X_{2}\sim -i\varepsilon  \)~~
~~ ~~~ ~~ ~~~ ~~ ~ ~~~~~ ~~~~~ ~~~ ~~ ~ ~~~
~~ ~~~ ~~ ~~ ~(4.24)

Now 

\( i \)G\( _{00} \)\( = \)\( \frac{1}{Y}=\frac{(X+i\varepsilon )}{(X-X_{1})(X-X_{2})} \)=\( \frac{X_{2}+i\varepsilon }{X_{2}-X_{1}}\frac{1}{X-X_{2}}+ \)\( \frac{X_{1}+i\varepsilon }{X_{1}-X_{2}}\frac{1}{X-X_{1}} \)~~~
~~ ~~~ ~~ ~~~ ~~ ~~~ ~~ ~~ ~(4.25)

The right hand side now shows, in the complex \( k_{0} \)-plane, 4 poles in
stead of 2: those at \( k_{0}^{2}= \) \( \frac{|\mathbf{k}|^{2}}{\theta ^{2}}+\frac{X_{1}}{\theta } \)
and \( k_{0}^{2}= \) \( \frac{|\mathbf{k}|^{2}}{\theta ^{2}}+X_{2} \).We also
note that \( X_{1,2} \) and hence the residues at these poles depend sensitively
on relative magnitude of \( \epsilon  \) compared to that of \( \delta  \)\( \theta  \).
We note in particular that for \( \epsilon  \) sufficiently large, \( X_{1}\sim -i\varepsilon /\theta  \)
and \( X_{2}\sim -i\varepsilon  \), so that the first term in (4.25) drops
out {[}residue tends to zero{]} and we are left with only two (usual) poles.
In this case, the residue at \( X=X_{1} \) also tends to one. This is what
one normally has. On the other hand, for sufficiently small \( \epsilon  \)
{[}and this limit is the one that interests us {]} we have \( |X \)\( _{1,2}|>>\varepsilon  \)
and \( X_{1}+X_{2}\simeq O(\varepsilon ) \); so that \( X_{1}-X_{2}=2X_{1}+O(\varepsilon ) \).
Thus the coefficients \( \frac{X_{2}+i\varepsilon }{X_{2}-X_{1}} \) \( \simeq  \)\( \frac{X_{1}+i\varepsilon }{X_{1}-X_{2}} \)\( \simeq  \)\( \frac{1}{2} \)
and all the 4 poles appear with equal weightage. For intermediate values of
\( \delta  \)\( \theta  \)/\( \epsilon  \) there are still 4 poles but the
residues vary continuously between the two extremes. Now, consider the contribution
for the coordinate space propagator 

G\( _{00}(x-y) \) \( \equiv  \)\( \int \frac{d^{4}k}{(2\pi )^{4}}\exp [-ik.(x-y)]G_{00}(k) \) 

and to be concrete,  let us consider the case \( x_{0}-y_{0}>0 \). To perform
the \( k_{0} \)-integration, we close the contour below. Then only the poles
in the lower quadrant will contribute to the integral.The residue at these poles
is variable as \( \epsilon  \) is varied. In fact, for \( \epsilon  \) >\textcompwordmark{}>
\( \delta  \)\( \theta  \), we have the positive energy pole contributing
as expected of a causal prescription, whereas for \( \epsilon  \) <\textcompwordmark{}<
\( \delta  \)\( \theta  \) both the positive and the negative energy poles
contribute equally. For intermediate values of \( \epsilon  \),  we have the
two poles contributing to a variable degree. 

After this demonstration of how delicately the \( \epsilon  \)-term is affected
by an infinitesimal change in the interpolating parameter \( \theta  \) {[}\( 1>\theta >0 \){]},
we shall make several observations:

(a) The (0,0) component of the propagator we have considered for simplicity
propogates an unphysical degree of freedom and as such, we may not find the
above situation undesirable \emph{purely} from a physical point of view. We
shall soon rectify this by also considering the spatial components and reaching
a similar conclusion.

(b) Nonetheless, we have demonstrated how unstable the ansatz of the \( \epsilon  \)-term
used in reference {[}7{]} is with respect to small variations of the interpolating
parameter \( \theta  \). It is only in an infinitesimal neighborhood of \( \theta  \)=1
{[}Feynman gauge{]}, where \emph{we know that the \( \epsilon  \)-term is correct,}
that the instability is absent. 

(c) This strongly suggests that we cannot use such an \( \epsilon  \)-term
if we are to preserve gauge-independence. 

(d) We noted that for a sufficiently small \( \epsilon  \), there is a \emph{apparently
discontinuous behavior} in the propagator as \( \theta  \) is varied by a small
amount. A propagator having contribution from positive energy pole suddenly
goes into one having equal contribution from positive and negative energy poles!
This looks suspicious as one may naively expect that as the formulation seems
continuous in \( \theta  \),  such discontinuity cannot appear. Here, we note
the existence of a ratio \( \delta  \)\( \theta  \)/\( \epsilon  \), which
can vary widely, and on which the modified \( \epsilon  \)-term sensitively
depends near a pole. Thus, the effect of the net \( \epsilon  \)-term is actually
continuous in \( \theta  \), but the scale of variation is too small \( \sim  \)
\( \epsilon  \). As \( \delta  \)\( \theta  \) <\textcompwordmark{}< \( \epsilon  \),
we in fact recover the old causal propagator; but for \( \delta \theta >>\varepsilon  \),
the propagator looks completely different as the residues at various poles depend
sensitively on \( \delta  \)\( \theta  \)/\( \epsilon  \)! 

(e)When it comes to choosing the \( \epsilon  \)-term for another type of gauge,
 a procedure was proposed in references {[}8,1{]} to do this and was followed
up in {[}9,10{]}. It was by transforming the Lorentz gauge path-integral \emph{with}
the correct \( \epsilon  \)-term, into a path-integral of desired class of
gauges. It is done with the help of the finite-field dependent BRS (FFBRS) transformation
developed for the appropriate pair of gauges {[}8{]}. This procedure, starting
from the Lorentz-type gauge, in effect, gathers along (integrates) the change
in the \( \epsilon  \)-term as a parameter such as \( \theta  \) is varied.
In this process, we would, thus, not find ourselves with the naive \( \epsilon  \)-term
for a \( \theta  \) < 1 {[} i.e. away from the Feynman gauge{]} in the first
place. The net result for Green's functions is accessible in a mathematically
tractable form {[}9,10{]}.

Having demonstrated the phenomenon with the help of the (0,0) component for
simplicity, we shall now proceed with the spatial components. The discussion
proceeds much the same way;  hence we shall give only the main steps. For spatial
components the propagator is 

\( G_{ij}=i\{Z+\varepsilon \delta \theta Z^{1}\}^{-1}_{ij} \) 

We note the values:

\( Z_{ij} \) = \( (k^{2}+i\varepsilon )[\delta _{ij}+k_{i}k_{j}A] \);  \( A=\frac{(1-\frac{1}{\theta ^{2}})}{k^{2}+i\varepsilon } \)

\( \varepsilon \delta \theta Z^{1}_{ij}= \) \( (k^{2}+i\varepsilon )k_{i}k_{j}\delta A \);
 \( \delta A= \)\( \frac{2i\varepsilon \delta \theta (1-\theta )}{\theta ^{3}[\theta k_{0}^{2}-\frac{1}{\theta }|\mathbf{k}|^{2}+i\varepsilon ](k^{2}+i\varepsilon )} \)

\( A'\equiv A+\delta A \)

\( G_{ij} \) is then given by

\( G_{ij}=i\{Z+\varepsilon \delta \theta Z^{1}\}^{-1}_{ij}=\frac{i}{k^{2}+i\varepsilon }\left( \delta _{ij}-k_{i}k_{j}\frac{A'}{1+|\mathbf{k}|^{2}A'}\right)  \)

It is the denominator of \( k_{i}k_{j} \) that has nontrivial structure. For
one of the terms, it is given by

\( (k^{2}+i\varepsilon )(1+|\mathbf{k}|^{2}A')=k^{2}+i\varepsilon +|\mathbf{k}|^{2}(1-\frac{1}{\theta ^{2}})+ \)\( \frac{2i\varepsilon |\mathbf{k}|^{2}\delta \theta (1-\theta )}{\theta ^{3}[\theta k_{0}^{2}-\frac{1}{\theta }|\mathbf{k}|^{2}+i\varepsilon ]} \)

~ ~~ ~~ ~~~ ~~ ~\( =\frac{X}{\theta }+i\varepsilon +\frac{2i\varepsilon \delta \theta (1-\theta )|\mathbf{k}|^{2}}{\theta ^{3}(X+i\varepsilon )} \)

Vanishing of this denominator implies

\( X^{2}+i\varepsilon (1+\theta )X-\varepsilon ^{2}\theta +\frac{2i\varepsilon \delta \theta (1-\theta )|\mathbf{k}|^{2}}{\theta ^{2}}=0 \)

The solutions to this equation are verified to have very similar properties
as those for (4.20): In fact, for \( \epsilon  \) sufficiently small,

\( X_{1}\simeq -X_{2}\simeq \sqrt{-2i\varepsilon \delta \theta (1-\theta )}\frac{|\mathbf{k}|}{\theta } \)\( \equiv  \)\( \left( 1-i\right) \frac{2|\mathbf{k}|}{\theta }\sqrt{\varepsilon \delta \theta (1-\theta )} \) 

which is >\textcompwordmark{}>\( \epsilon  \) in magnitude.

and for \( \epsilon  \) >\textcompwordmark{}> \( \delta  \)\( \theta  \),
 we have 

\( X_{1}\sim -i\varepsilon \theta  \) and \( X_{2}\sim -i\varepsilon  \)

In view of this exactly analogous conclusions follow for the spatial components
also.

\section{A SIMPLE EXAMPLE OF QED}

In the last section, we had shown that if one were to use the term \( -i\varepsilon \int d^{4}x\left( \frac{1}{2}A^{2}-\overline{c}c\right)  \)
in the path- integral for the interpolating gauges, we develop a severe pathological
behavior. As mentioned in Section 3 {[}see before Eq.(3.2){]}, the \( \epsilon  \)-term
is not fixed in all details. A question, then, naturally arises if there is
no simple modification of this epsilon-term that will do the job. An essential
requirement for such a term to lead to no residual contributions of the kind
(3.15)(at least to the lowest order) has been the invariance of such a term
under the abelian part of BRS.

In the case of the Doust interpolating scheme {[}7{]}, a term

\( -i\varepsilon \theta \int d^{4}x\left( \frac{1}{2}A_{0}^{2}-\frac{1}{2\theta ^{2}}\mathbf{A}^{2}-\frac{1}{\theta }\overline{c}c\right)  \)
~ ~~ ~~~ ~ ~~ ~~ ~~ ~~~ ~~ ~ ~~ ~~ ~~ ~~~ ~~ ~~
~~(5.1)

{[}or any term proportional to it (by a positive number) {]} does in fact satisfy
the condition for invariance under the abelian part of BRS\footnote{%
Here, we are refering to the invariance in the lowest order.
} :

\( \delta A_{\mu }(x)= \) \( \partial _{\mu } \)c(x) \( \delta  \)\( \Lambda  \)

\( \delta c^{\alpha }(x)= \)0

\( \delta \overline{c}^{\alpha }(x)= \) \( [\theta \partial ^{0}A_{0}-\frac{1}{\theta } \)\( \partial  \)\( _{i}A_{i}] \)
\( \delta  \)\( \Lambda  \) ~ ~~ ~ ~~ ~~~ ~ ~~ ~~ ~~ ~ ~
~~ ~ ~~ ~~ ~~ ~~~ ~~ ~~ ~~(5.2)

Thus a question naturally arises whether such a simple modification may not
always exist and whether (or not) it will remove the problem at hand.

Another question the treatment of the last section will perhaps raise is whether
any delicate approximations have been made in the process there that could lead
to such an unusual result.

With these questions in mind, we shall work out a simple example in the context
of QED. We shall be able to work out this example exactly without the need for
any simplifying approximations . We shall show that \emph{even with the \( \epsilon  \)-term
(5 .1) finely tuned to cancel the leading term in (3 .15),  there is a residue
that again leads to the same phenomenon.}

Being an abelian theory in linear gauges, we can dispense ourselves with ghosts
and confine ourselves only with the gauge transformations .

Consider, thus, the expectation value of a gauge-invariant operator \( O[A,\psi ,\overline{\psi }]\equiv O[\phi ] \):

\( <O[\phi ]>=\frac{1}{W}<<O[\phi ]>> \)~~ ~ ~~ ~~~ ~ ~~ ~~ ~~
~~~ ~~~ ~ ~~ ~~ ~~ ~~~ ~~ ~~ ~~ ~~ ~~(5.3a)

with,

\( <<O[\phi ]>>\equiv \int D\phi O[\phi ]\exp \{iS_{0}[A,\psi ,\overline{\psi }]-\frac{i}{2}\int d^{4}xF[A,\theta ]^{2}+ \)
an \( \epsilon  \)-term \}~~ ~~~~ ~~(5.3b)

and 

\( W\equiv <<1>> \). 

In the path integral, we can perform any kind of a field transformation, in
particular, any gauge transformation, without altering its value. We shall consider
the following gauge transformation\footnote{%
We shall use below the summation integration convention of DeWitt for the sake
of compactness. See e.g.reference {[}17{]} for its usage. Here, we note for
brevity a few things:In \( A_{i} \) , \( i \) refers to a space-time lable
\( x_{i} \) , as well as a Lorentz index \( \mu  \) . Compact indices are
raised and lowered by \( g^{ij} \)\textasciitilde{} \( g^{\mu \nu }\delta ^{4}(x_{i}-x_{j}) \).
In \( F^{\alpha } \), \( \alpha  \) refers to a space-time label \( x_{\alpha } \).
\( \delta  \)\( ^{\alpha \beta } \)\textasciitilde{} \( \delta  \)\( ^{4}(x_{\alpha }-x_{\beta }) \)
raises these indices. \( F^{\alpha }F_{\alpha } \) stands for \( \int d^{4}x \)\( F(x)^{2} \).
} , not necessarily infinitesimal,

\( A_{i} \) \( \equiv  \)\( A'_{i}+ \)\( \partial  \)\( _{i}^{\alpha }\left( M_{0}-i\varepsilon \right) ^{-1}_{\alpha \beta }\Delta F^{\beta }[A'] \)
~~ ~ ~~ ~~~ ~ ~~ ~~ ~~ ~~~~ ~~~~ ~ ~~ ~~ ~~ ~~~
~~ ~~ ~~(5.4)

with

\( M^{\alpha \beta }_{0}=\frac{\delta F^{\alpha }}{\delta A_{i}}\partial _{i}^{\beta } \)~~
~ ~~ ~~~ ~ ~~ ~~ ~~ ~~~ ~~ ~~~~ ~~~~~ ~ ~~ ~~
~~ ~~~ ~~ ~~ ~~(5.5)

(\( M_{0} \) is \( A \)-independent) and an appropriate gauge transformation
on \( \psi  \) and \( \overline{\psi } \) {[}with a field-dependent gauge
parameter{]}. Here, we note the value \( \partial ^{\alpha }_{i}=\partial ^{^{x_{i}}}_{\mu }\delta ^{4}(x_{i}-x_{\alpha }) \)
(which is a \emph{number)} when \( \alpha  \) stands for \( (x_{\alpha }) \)
and \( i \) for \( (\mu ,x_{i}) \). We parametrize the \( \epsilon  \)-term
in (5.3b) as 

\( \frac{1}{2}\varepsilon a^{ij}A_{i}A_{j} \)~~ ~ ~~ ~~~ ~ ~~ ~~
~~ ~~~ ~~~ ~ ~~~~~ ~~~ ~ ~~ ~~ ~~ ~~~ ~~ ~~ ~~(5.6)

with \( a_{ij}=a_{ji} \) and \( a_{ij} \) being local, i.e. proportional to
\( \delta  \)\( ^{4}(x_{i}-x_{j}) \) or its derivatives.

For a \emph{linear} \( F^{\alpha } \) and \( \delta F^{\alpha } \), the transformation
(5.4) is linear in \( A \). The super-determinant for the transformation is
a constant i.e. independent of fields ( \( D\psi D\overline{\psi } \) being
invariant) and thus does not contribute to the ratio (5.3a).

Under the gauge transformation, \( S_{0} \) is invariant. Under this field
transformation (we recall that \( F \) is linear in \( A \)),

\( F^{\alpha }[A]=F^{\alpha }[A']+\frac{\delta F^{\alpha }}{\delta A_{i}}\partial ^{\beta }_{i} \)\( \left( M_{0}-i\varepsilon \right) _{\beta \gamma }^{-1}\Delta F^{\gamma }[A'] \)

\( \quad \quad \quad  \) =  \( F^{\alpha }[A']+ \)\( M^{\alpha \beta }_{0}\left( M_{0}-i\varepsilon \right) _{\beta \gamma }^{-1}\Delta F^{\gamma }[A'] \)

\( \quad \quad \quad  \) = \( F^{\alpha }[A']+ \)\( \Delta F^{\alpha }[A']+i\varepsilon [\left( M_{0}-i\varepsilon \right) ^{-1}]^{\alpha }_{\gamma }\Delta F^{\gamma }[A'] \)~~
~ ~~ ~~~ ~ ~~ ~~ ~~ ~~~ ~~ ~~ ~~(5.7)

The gauge fixing term goes into

\( -\frac{i}{2}F^{\alpha }[A]F_{\alpha }[A]\equiv  \)\( -\frac{i}{2}F^{\alpha ^{2}}[A]= \)\( -\frac{i}{2}\{F^{\alpha }[A']+\Delta F^{\alpha }[A']\}^{2}+\varepsilon F^{\alpha }[A']\left( M_{0}-i\varepsilon \right) _{\alpha \beta }^{-1}\Delta F^{\beta }[A']+ \)

~ ~~~ ~ ~~ ~\( \varepsilon \Delta F^{\alpha }[A']\left( M_{0}-i\varepsilon \right) _{\alpha \beta }^{-1}\Delta F^{\beta }[A']+ \)\( \frac{i}{2}\varepsilon ^{2}\{ \)\( \left( M_{0}-i\varepsilon \right) _{\alpha \beta }^{-1}\Delta F^{\beta }[A'] \)\}\( ^{2} \)~~~
~ ~~ ~~ ~~ ~~~ ~~ ~~ ~~(5.8)

and the \( \epsilon  \)-term goes into

\( \frac{1}{2}\varepsilon a^{ij}A_{i}A_{j}= \)\( \frac{1}{2}\varepsilon a^{ij}A'_{i}A'_{j} \)+\( \varepsilon a_{ij}A'_{i} \)\( \partial  \)\( _{j}^{\beta }\left( M_{0}-i\varepsilon \right) ^{-1}_{\beta \gamma }\Delta F^{\gamma }[A'] \)+ 

~ ~~ ~~ ~~ \( \frac{1}{2}\varepsilon a^{ij} \)\( \partial  \)\( _{i}^{\alpha }\left( M_{0}-i\varepsilon \right) ^{-1}_{\alpha \beta }\Delta F^{\beta }[A'] \)\( \partial  \)\( _{j}^{\delta }\left( M_{0}-i\varepsilon \right) ^{-1}_{\delta \gamma }\Delta F^{\gamma }[A'] \)~~~~~
~~ ~~~ ~~ ~~~ ~~ ~~ ~~(5.9)

We note that the terms in (5.8) and (5.9) of the leading order in \( \Delta F \)
together are

\( \varepsilon F^{\alpha }[A']\left( M_{0}-i\varepsilon \right) _{\alpha \beta }^{-1}\Delta F^{\beta }[A']+ \)\( \varepsilon a^{ij}A'_{i} \)\( \partial  \)\( _{j}^{\beta }\left( M_{0}-i\varepsilon \right) ^{-1}_{\beta \gamma }\Delta F^{\gamma }[A'] \)~
~ ~~ ~~ ~~ ~~~ ~~~~~ ~~ ~~ ~~(5.10)

It is the equivalent of these terms, that were discussed in their effects, in
section 4. These terms together can cancel out, for an arbitrary \( \Delta F \),
if and only if for a given \( F \), a choice of \( a_{ij} \) can be made such
that\footnote{%
This relation based on the above cancellation is specific to the gauge transformation
(5.4).
}

\( F^{\alpha }= \) \( -a^{ij}\partial ^{\alpha }_{j}A_{i} \)~ ~ ~~ ~~
~~ ~~~ ~~ ~~ ~~~ ~~ ~~ ~~ ~~~ ~~ ~~~~ ~~~~~~
~~ ~~ ~~~ ~~ ~~ ~~(5.11)

Now if the \( \epsilon  \)-term is not or cannot be chosen consistent with
(5.11), then the entire set of observations in section 4 (based upon this term)
hold. Presently, we want to study the possibility that the two terms in (5.10)
do cancel. Later we shall make observations on when this is /is not possible.

Assuming that (5.11) holds, we have,

\( M^{\alpha \beta }_{0}=\frac{\delta F^{\alpha }}{\delta A_{i}}\partial _{i}^{\beta }=-a^{ij}\partial _{j}^{\alpha }\partial _{i}^{\beta }=M^{\beta \alpha }_{0} \)
~ ~~ ~~ ~~ ~~~ ~~ ~~ ~ ~~ ~~ ~~~ ~~ ~~~ ~ ~~ ~~
~~~ ~~ ~~ ~~(5.12)

Using this we can simplify the last term in (5.9) as

\( \frac{1}{2}\varepsilon a^{ij} \)\( \partial  \)\( _{i}^{\alpha }\left( M_{0}-i\varepsilon \right) ^{-1}_{\alpha \beta }\Delta F^{\beta }[A'] \)\( \partial  \)\( _{j}^{\delta }\left( M_{0}-i\varepsilon \right) ^{-1}_{\delta \gamma }\Delta F^{\gamma }[A'] \) 

\( =-\frac{1}{2}\varepsilon  \)\( M^{\alpha \delta }_{0}\left( M_{0}-i\varepsilon \right) ^{-1}_{\alpha \beta }\Delta F^{\beta }[A'] \)\( \left( M_{0}-i\varepsilon \right) ^{-1}_{\delta \gamma }\Delta F^{\gamma }[A'] \)

\( =-\frac{1}{2}\varepsilon  \)\( \left( M_{0}-i\varepsilon \right) ^{-1}_{\alpha \beta }\Delta F^{\beta }[A'] \)\( \Delta F^{\alpha }[A'] \)\( -\frac{i}{2}\varepsilon ^{2} \)\( \left( M_{0}-i\varepsilon \right) ^{-1}_{\alpha \beta }\Delta F^{\beta }[A'] \)\{\( \left( M_{0}-i\varepsilon \right) ^{-1}\}_{\gamma }^{\alpha }\Delta F^{\gamma }[A'] \)~~
~~ ~~ ~~~ ~~ ~~ ~~(5.13)

we collect the remaining terms in (5.8) and (5.9) together and after some cancellations
and simplifications, we find that the gauge-fixing and the \( \epsilon  \)-terms
together become,

\( \frac{-i}{2}F^{\alpha ^{2}} \)\( -\frac{1}{2}\varepsilon a^{ij}A_{i}A_{j} \)
= \( \frac{-i}{2}\{F[A']+\Delta F[A']\}^{2}+\frac{1}{2}\varepsilon a^{ij}A'_{i}A'_{j}+ \)
\( \frac{1}{2}\varepsilon \Delta F^{\alpha }[A']\left( M_{0}-i\varepsilon \right) _{\alpha \beta }^{-1}\Delta F^{\beta }[A'] \)~
~~ ~~ ~~~ ~~ ~~ ~~(5.14)

We thus find that 

\( <<O[\phi ]>>\equiv \int D\phi O[\phi ]\exp \{iS_{0}[A,\psi ,\overline{\psi }]-\frac{i}{2}\int d^{4}xF[A,\theta ]^{2} \)\( +\frac{1}{2}\varepsilon a^{ij}A_{i}A_{j} \)
\}

\( =\int D\phi 'O[\phi ']\exp \{iS_{0}[A',\psi ',\overline{\psi }']-\frac{i}{2}\int d^{4}x\{F[A',\theta ]+\Delta F[A']\}^{2} \)\( +\frac{1}{2}\varepsilon a^{ij}A'_{i}A'_{j} \)
+\( \frac{1}{2}\varepsilon \Delta F^{\alpha }\left( M_{0}-i\varepsilon \right) _{\alpha \beta }^{-1}\Delta F^{\beta } \)\}

~ ~~ ~~~~ ~~ ~~~ ~~ ~~ ~~ ~~~~ ~~ ~~~ ~~ ~~ ~~~
~~ ~~~ ~~ ~~~~ ~~~ ~~ ~~ ~~~~ ~~ ~~~ ~~ ~~ ~~(5.15)

We note that we have derived the above result exactly and not necessarily for
an infinitesimal \( \Delta F \). Here, we note the presence of the last term
in the exponent that has precisely similar effects as the last term in (4.6)
that were elaborately discussed in Section 4. To see this, consider {[}we shall
drop primes on \( A' \) henceforth{]}

\( F[\theta ,A]= \)\( [\theta \partial ^{0}A_{0}-\frac{1}{\theta } \)\( \partial  \)\( _{i}A_{i}] \)
; \( \Delta F\equiv F[\theta ',A]-F[\theta ,A]=(\theta '-\theta )\left( \partial _{0}A_{0}+\frac{1}{\theta '\theta }\partial _{i}A_{i}\right)  \)~~
~~ ~~(5.16)

Here, we need not yet require \( (\theta '-\theta ) \) an infinitesimal. We
now focus our attention on the part of the action quadratic in \( A \).  It
is

\( iS_{0}[A]-\frac{i}{2}F^{\alpha }[\theta ',A]^{2}+\frac{1}{2}\varepsilon \theta [A^{2}_{0}-\frac{1}{\theta ^{2}}\mathbf{A}^{2}]+\frac{1}{2}\varepsilon \Delta F^{\alpha }\left( M_{0}-i\varepsilon \right) _{\alpha \beta }^{-1}\Delta F^{\beta } \)
~~~~ ~~~ ~~ ~~ ~~~~ ~~ ~~~ ~~ ~~ ~~(5.17)

This in particular leads to 

\( Z'_{00}= \)|\textbf{k|\( ^{2} \)\( - \)\( \theta ' \)\( ^{2} \)\( k_{0}^{2} \)\( - \)\( i\varepsilon \theta  \)}
\( \left( 1-\frac{k^{2}_{0}(\theta '-\theta )^{2}/\theta }{\theta k^{2}_{0}-\frac{1}{\theta }|\mathbf{k}|^{2}+i\varepsilon }\right)  \)

We note that the additional term is of \( O[(\theta '-\theta )^{2}] \). We
can now carry out the entire analysis in parallel as in section 4, and reach
similar conclusions.

Finally, we comment that a choice of the \( \epsilon  \)-term satisfying (5.11)
{[}where \( a_{ij} \) is symmetric{]} may not exist for a given \( F^{\alpha } \).
If this is not possible, the analysis of section 4 is sufficient to draw our
conclusions. We have seen explicitly that for the Doust {[}7{]} interpolating
gauge, \( a_{ij} \) do exist, so that the analysis of this section is required
if this choice is made. For the gauges \( (P\partial )A=0 \) {[}18{]}, one
can find \( a_{ij} \) if P\( _{\mu \nu } \) {[} with both lower or both upper
indices{]} is symmetric. Thus, one cannot choose \( a_{ij} \) for the light
front gauge fixing {[}18{]} \( (\partial _{0}+\partial _{3})(-A_{0}+A_{3})=F=0 \). 

The important point made in this section is that the variation of a parameter
in interpolating gauges is associated with gauge transformations such as (5.4),
which as \( \epsilon  \)\( \rightarrow 0 \), become singular; thus, suggesting
that care is needed in operations involving such parameter variations.

\section{CONCLUSIONS}

We shall now summarize the conclusions for the work presented. We emphasized
the importance of the correct \( \epsilon  \)-term in the discussion of gauge-independence.
We explicitly demonstrated that to keep the gauge-independence of , say , the
vacuum expectation value of a gauge-invariant operator, it is generally necessary
to vary the \( \epsilon  \)-term appropriately with parameters in \( F[A,\alpha ] \).
We explained, how in the context of the family of Lorentz gauges, this alteration
in the \( \epsilon  \)-term does not alter the causal nature of the prescription.
We explained why the \( \epsilon  \)-term need not be paid attention to in
the discussion of gauge-independence for the class of the Lorentz-type gauges
as \( \epsilon  \)\( \rightarrow 0 \) and emphasized that this was an exception
rather than a rule. We illustrated the effect of the modification of the \( \epsilon  \)-term
with a variation of parameter by applying it to the interpolating gauges used
by Doust {[}7{]} to connect the Feynman and the Coulomb gauges together with
a fixed \( \epsilon  \)-term added to the action for all \( \theta  \): 1\( \geq \theta \geq 0 \).
We demonstrated that as \( \theta  \) is varied, gauge-independence cannot
be preserved unless the \( \epsilon  \)-term is suitably varied and thus that
such an interpolating gauge does not really gauge-invariantly interpolate between
the Lorentz gauge and the Coulomb gauge. What is worse is that the small variation
in \( \epsilon  \)-term with \( \theta  \) has a catastrophic effect: it the
gauge boson propagator structure from the causal one to a mixed one even for
a small change \( \delta  \)\( \theta  \). We further demonstrated that there
was no modification of the \( \epsilon  \)-term that would allow us an escape.

\textbf{\emph{ACKNOWLEDGEMENTS}}

I would like to acknowledge financial support from Department of Science and
Technology,  Government of India in the form of a grant for the project DST-PHY-19990170.

\section{Appendix A}

In this appendix, we shall discuss the behavior of the sign of the imaginary
part of Y {[}See Eq.(4.18){]} near the point where the real part vanishes. This
gives an alternate semi-quantitative view of what happens. 

Consider an example of a contribution to the Green's function ,

\( \int  \)\( dk_{0} \) \( \frac{\exp \{-ik_{0}t\}}{k_{0}-\omega } \)~~
~~ ~~~~ ~~ ~~~~ ~~ ~~ ~~~~ ~~~ ~~ ~~ ~~~~ ~~
~~~ ~~ ~~ ~~(A.1)

This is ill-defined and the definition is made by specifying how the singularity
at \( k_{0}=\omega  \) is treated. A prescription such as 

\( \int  \)\( dk_{0} \) \( \frac{\exp \{-ik_{0}t\}}{k_{0}-\omega +i\varepsilon } \)
implies that the contour in (A.1) near \( k_{0}=\omega  \) is to be taken above
the real axis. We can of course distort the contour as we like \emph{elsewhere}
and not have the value of the integral changed in view of the analyticity of
the integrand.Since the specification of the location of the contour \emph{near}
\( k_{0}=\omega  \) is all that counts, we could as well redefine the Green's
function as

\( \int  \)\( dk_{0} \) \( \frac{\exp \{-ik_{0}t\}}{k_{0}-\omega +i\varepsilon (k_{0})} \)

where \( \epsilon  \)(k\( _{0}) \) is any analytic function.{[}Example:\( \epsilon  \)(k\( _{0}) \)
= \( \epsilon  \)(\( k_{0} \)\( ^{2} \)+\( m^{2} \)){]}. To reemphasize,
it is the \emph{sign} of the \( \epsilon  \)-term \emph{near} the singularity
of the real part that matters as \( \epsilon  \)\( \rightarrow  \)0.

Now, at \( \delta  \)\( \theta  \)\( =0, \) the sign of the imaginary part
at the location of the \( ReY=0, \) i.e. \( X=0 \) or \( k_{0}=\pm \frac{1}{\theta }|\mathbf{k}| \)
is positive and this determines on which side of the real axis the poles are
and this in turn determines how they contribute say to the coordinate propagator.
We would like to study how this is affected when \( \delta  \)\( \theta  \)
\( \neq  \)0. Ordinarily, i.e., if the hypothesis of {}`` causal{}'' prescription
in (4.1) is in fact correct, we would by continuity expect nothing unusual.
Now,  from (4.18),

\( ReY=\theta X-\frac{2\delta \theta \varepsilon ^{2}(1-\theta )k_{0}^{2}}{X^{2}+\varepsilon ^{2}} \)
;  \( ImY=\varepsilon -\frac{2\delta \theta \varepsilon (1-\theta )k_{0}^{2}X}{X^{2}+\varepsilon ^{2}} \)

When \( ReY=0, \) i.e.\footnote{%
In the following we drop the prime on \( \theta  \) for convenience.
},

\( \theta X=\frac{2\delta \theta \varepsilon ^{2}(1-\theta )k_{0}^{2}}{X^{2}+\varepsilon ^{2}} \)

we have,

\( \frac{\theta X^{2}}{\varepsilon ^{2}} \)\( =\frac{2\delta \theta (1-\theta )k_{0}^{2}X}{X^{2}+\varepsilon ^{2}} \)

which in turn leads to 

\( ImY=\varepsilon \left( 1-\frac{\theta X^{2}}{\varepsilon ^{2}}\right) =-\varepsilon \theta \left( 1+\frac{X^{2}}{\varepsilon ^{2}}\right) +\varepsilon (1+\theta ) \) 

Further, \( ReY=0, \)implies that 

\( \theta X(X^{2}+\varepsilon ^{2})+2\delta \theta \varepsilon ^{2}(1-\theta )k_{0}^{2}=0 \)

so that at \( \varepsilon =0 \), we have \( X=0. \)

We further note,

\( \theta X=\frac{2\delta \theta (1-\theta )k_{0}^{2}}{\frac{X^{2}}{\varepsilon ^{2}}+1} \)

Now , as \( \epsilon  \) \( \rightarrow  \) 0, we know that \( X \)\( \rightarrow  \)
0.This implies that \( \{1+\frac{X^{2}}{\varepsilon ^{2}}\} \)\( \rightarrow  \)
\( \infty  \) .Thus, for a fixed value of \( \delta  \)\( \theta  \),  we
always have a sufficiently small \( \epsilon  \) such that {[}for \( \theta  \)
> 0{]},

\( ImY= \)\( \epsilon \left( 1+\theta -\theta \left[ 1+\frac{X^{2}}{\varepsilon ^{2}}\right] \right) <0 \) 

i.e. the sign of the imaginary part changes.

\end{document}